\documentclass[preprint]{aastex}
\usepackage{graphicx}

\begin{document}

\title{MASS OUTFLOW FROM RED GIANT STARS IN M13, M15, AND M92}

\author{Sz. M{\'e}sz{\'a}ros\altaffilmark{1,3}, E. H. Avrett\altaffilmark{2,4}, 
and A. K. Dupree\altaffilmark{2,5}}

\altaffiltext{1}{Department of Optics and Quantum Electronics, University of Szeged, 
6701 Szeged, Hungary}
\altaffiltext{2}{Harvard-Smithsonian Center for Astrophysics, Cambridge, MA 02138}
\altaffiltext{3}{e-mail address: meszi@physx.u-szeged.hu}
\altaffiltext{4}{e-mail address: eavrett@cfa.harvard.edu}
\altaffiltext{5}{e-mail address: dupree@cfa.harvard.edu}

\begin{abstract}
Chromospheric model calculations of the H$\alpha$ line for selected
red giant branch (RGB) and asymptotic giant branch (AGB)  
stars in the globular clusters M13, 
M15, and M92 are constructed to derive mass loss rates. The model
spectra are compared to the observations 
obtained with the Hectochelle on the MMT telescope. These stars show strong 
H$\alpha$ emissions and blue-shifted H$\alpha$ cores signaling that
mass outflow is present in all stars. Outflow velocities of 
3$-$19~km~s$^{-1}$, larger than indicated by H$\alpha$ profiles, are needed 
in the upper chromosphere to achieve good agreement between the model
spectra and the observations. The resulting
mass loss rates range from 0.6$\times$10$^{-9}$ to 5$ \times
$10$^{-9}$~M$_{\odot}$~yr$^{-1}$, which are 
about an order of magnitude lower than predicted from ``Reimers' law"
or inferred from the infrared excess of similar stars. 
The mass loss rate increases slightly with luminosity 
and with decreasing effective temperature. Stars in the more metal-rich M13 have 
higher mass loss rates by a factor of $\sim$2  than in the metal-poor
clusters M15 and M92. 
A fit to the mass loss rates is given by: 
$\dot M$[M$_{\odot}$~yr$^{-1}$] = 0.092 $\times$
L$^{0.16}\times$ T$_{\rm eff}^{-2.02}\times $A$^{0.37}$ where A=10$^{[Fe/H]}$.
Multiple observations of stars revealed one object in M15, K757, in which the mass 
outflow increased by a factor of 6 between two observations
separated by 18 months. Other stars showed changes in mass loss rate by a factor of 1.5 or
less. 

\end{abstract}


\section{Introduction}

Stellar evolution theory predicts that low-mass Population II stars
ascending the red giant branch (RGB) for the first time must lose
mass \citep{renzini01, sweigart02}. \citet{iben1970} conjectured that
mass loss on the RGB may increase with metallicity in order to account
for colors on the horizontal branch. Direct observations of the
ongoing mass loss process in globular clusters only became possible in the
past decade using high resolution spectroscopy and infrared imaging 
from space. For stellar evolution calculations, the mass loss rate from
late-type giants is frequently described 
by ``Reimers' law'' \citep{reimers01, reimers02} given as
$\dot{M}[M_\sun yr^{-1}]=\eta \times L_{*} \times R_{*}/M_{*}$, 
where $L_{*}, R_{*},$ and $M_{*}$ are the stellar luminosity, radius, and mass  in solar units, 
and $\eta$ is a fitting parameter equal to  $4 \times 10^{-13}$.  This
approximation is based on a handful of luminous Population I stars.  
\citet{schroder01} offered another semi-empirical relation for the mass
loss rate from cool stars by assuming a wave-driven wind and introducing
gravity and effective temperature into the formulation. They found
consistency with calculations of evolutionary models for abundances
as low as [Fe/H]=$-$1.27 although metallicity does not enter as a 
parameter in their formulation.

\citet{origlia02} identified dusty RGB stars in 47 Tuc from {\it Spitzer}
mid-IR photometry and derived an empirical (dust) mass loss law 
for these stars. Mass loss rates
derived from their observations suggested 
that the mass loss increases with luminosity, and is episodic since an
infrared excess is not found in all stars. \citet{boyer02} also detected a 
population of dusty red giants near the center of M15. The similarities in H$\alpha$ line profile 
characteristics between the \textit{Spitzer} sources and other red giants in M15 suggests the IR 
emission attributed to circumstellar dust must be produced by an episodic 
process \citep{meszaros02}. \citet{boyer01} observed dust production
possibly associated with 3 AGB 
stars in $\omega$~Cen with \textit{Spitzer}. They estimated 
a high mass loss rate for these AGB stars of
2.9$-$4.2$\times$10$^{-7}$ M$_{\odot}$~yr$^{-1}$; 
however, normal RGB stars in $\omega$~Cen do not appear to show strong 
mass loss as evidenced by the presence of dust. 

Indirect evidence of mass loss processes would be the detection of an intracluster medium.
Ionized intracluster gas was found in 47 Tucanae by measuring the radio dispersion 
of millisecond pulsars in the cluster \citep{freire01}, while
intracluster dust in M15 was first identified 
by \citet{evans01}. None of the detections reveal the  amount of intracluster material 
expected from the cluster giants \citep{barmby01,matsunaga2008}. 

Stars in several clusters have been examined with high resolution
spectroscopy  to search for  
signs of mass loss in their H$\alpha$ and 
\ion{Ca}{2}~K profiles. A detailed study carried out by
\citet{cacciari01} included 137 red giant 
stars in NGC~2808. Most of the stars brighter than log~$(L/L_{\odot})=2.5$ showed 
emission wings in H$\alpha$. The velocity shift of the H$\alpha$ line
core relative to the photosphere was  
less than $\approx -9$~km~s$^{-1}$. \citet{meszaros02, meszaros01} also found velocity shifts in the 
H$\alpha$ absorption line in M13, M15 and M92 down to a limiting
luminosity of log~$(L/L_{\odot})=2.0$. The line bisector 
showed increasing outflow velocities in the H$\alpha$ core up to 7$-$10~km~s$^{-1}$ above 
log~$(L/L_{\odot})\sim2.5$. The \ion{Ca}{2}~K line suggested higher
velocities and an accelerating outflow. The outflow velocities appear
to be independent of cluster metallicity. 

In order to evaluate the mass flow, detailed non-LTE modeling
with semi-empirical atmospheres is necessary to reproduce the optical
line profiles and infer the mass loss 
rates from the stars. Such non-LTE modeling was first carried out 
by \citet{dupree01}. They showed that the
emission wings of the H$\alpha$ line found in metal-deficient giant
stars can arise naturally from an extended, static chromosphere, and
emission asymmetry
and shifts in the H$\alpha$ core indicate mass loss. Spherical models with expanding
atmospheres suggested the mass loss rates are less than
2$\times$10$^{-9}$ M$_\sun$~yr$^{-1}$ a value which is less than predicted by the
Reimer's relationship. \citet{mcdonald01} calculated mass loss rates of two 
stars in M15 by modeling the H$\alpha$ and \ion{Ca}{2}~K lines with simple LTE approximations. 
They found mass loss rates of several times 10$^{-8}$ and 10$^{-7}
$M$_{\odot}$~yr$^{-1}$, but the use of LTE models for a chromosphere
can not be considered reliable. 
\citet{mauas01} computed semi-empirical H$\alpha$ and \ion{Ca}{2}~K
profiles for 5 
RGB stars in NGC~2808 including non-LTE effects in spherical coordinates. 
Their line profiles fit the 
observations when an outward velocity field is included in the model chromosphere, in agreement with 
previous calculations \citep{dupree01}. The derived mass loss rates
exhibited a large range around 10$^{-9}$~M$_{\odot}$~yr$^{-1}$. 
Outflow velocities from 10~km~s$^{-1}$ up to 80~km~s$^{-1}$ were
needed by \citet{mauas01} in order to match the observed line profiles. \citet{lyons01} discussed the 
H$\alpha$ and Na~I~D line profiles for RGB stars in M4, M13, M22, M55, and $\omega$~Cen.
The core shifts were less than 10~km~s$^{-1}$, much smaller than the escape velocity from the stellar 
atmosphere at $2 \ R_{*}$ ($\approx 50-70$~km~s$^{-1}$). \citet{dupree02} 
observed 2 RGB stars in NGC~6752 and found that the \ion{Ca}{2}~K and H$\alpha$ core shifts were
also low, less than 10~km~s$^{-1}$. However, asymmetries in the Mg~II
lines showed strong outflow velocities 
($\approx 150$~km~s$^{-1}$) in cluster giants and metal-poor field
stars \citep{dupree02, dupree2007, smith1998}. Also, high 
outflow velocities, (30$-$140~km~s$^{-1}$), were found in 
the He I $\lambda$10830 absorption line of metal-poor red giant stars
of which 6 are in M13 \citep{dupree1992, smith03, dupree2009}. 
Mg~II lines and He I $\lambda$10830 are formed higher in the 
atmosphere than H$\alpha$ and \ion{Ca}{2}~K, 
which suggests that the stellar wind becomes detectable near the top of the chromosphere. These outflow
velocities are frequently higher than the central escape velocities from globular
clusters, namely 20$-$70~km~s$^{-1}$ \citep{mclaughlin01}. 

In this paper, we select a sample of giant stars to model whose spectra have been 
obtained previously with Hectochelle \citep{meszaros02,meszaros01}. They span a 
factor of 5 in metallicity (from [Fe/H]=$-$1.54 
to $-$2.28)  and a factor of 6 in luminosity [from log~$(L/L_{\odot})=$2.57 to 3.38].  
Five stars have been observed more than once. Characteristics of the selected targets 
are described in Section 2. Section 3 contains the details of the non-LTE models in both the static 
and expanding versions. Section 4 compares the calculations with H$\alpha$ line profiles, and the 
construction of a mass loss relation and its dependence on temperature, luminosity, and abundance. 
Our Conclusions can be found in Section 5.

\section{Target Stars}

Observations of H$\alpha$ in a total of 297 red giant stars in M13,
M15, and M92 were obtained in 2005 May, 2006 May, and 
2006 October with the Hectochelle on the MMT \citep{meszaros02, meszaros01} 
with a spectral resolution of about 34,000. To investigate the
dependence of the mass loss rate on luminosity, temperature, and
metallicity, we chose RGB stars from each cluster that had clear H$\alpha$ emission  
and a range of at least a factor of two in luminosity. 
The sample of stars includes different intensity ratios of the
H$\alpha$ emission wings, B/R\footnote{B signifies 
the short-wavelength emission peak and R the long-wavelength emission
peak.}, and the bisector velocities varied 
($v_{\rm bis}$) from $-$0.7 to $-$8.9~km~s$^{-1}$. To monitor the mass loss changes in time, one star 
in M13 (L72), one in M92 (VII-18), and three stars in M15 (K341,
K757, and K969) were selected 
which have multiple observations. One star without H$\alpha$ emission was also selected
from each cluster to extend the sample to
lower luminosities and higher temperatures. Our previous study of M15 
\citep{meszaros02} found no signature
of different outflows or chromospheric structure between the `dusty'
stars identified by the \textit{Spitzer} Space Telescope \citep{boyer02} and normal
RGB objects. Modeling of the H$\alpha$ profile could reveal dynamical 
differences, if present. Two AGB stars in M15 with excess dust (K421,
K479) were selected for modeling to investigate whether the mass loss of these stars differs from the rest. 
The color magnitude diagram (CMD) of the 
cluster members and targets for modeling can be seen in Figure 1; target stars are listed in Table 1. 

A total of 15 stars was selected including from 4 to 6 in each 
of the 3 globular clusters; five stars had multiple spectra so that
we could estimate changes in the mass loss rate from the H$\alpha$ profiles. 
Unreddened colors for M13, M15, and M92 stars were calculated using
the foreground reddening and the apparent 
distance modulus from the catalog of \citet{harris01}.
The effective temperatures, bolometric corrections, and luminosities
were obtained from the V$-$K colors 
using the empirical calibrations by \citet{alonso01, alonso02} and the
cluster average metallicity \citep{harris01}
[Fe/H]=$-1.54$ for M13, [Fe/H]=$-2.26$ for M15, [Fe/H]=$-2.28$ for M92  [see 
\citet{meszaros02, meszaros01} for more details].

\section{The Models}

Our technique consists of constructing a static photosphere and
chromosphere model and then changing the
temperature at each depth until the H$\alpha$ profile depth, width, and 
emission strength and asymmetries are roughly consistent with observed 
profiles. At this point, the model
atmosphere and profile are calculated in spherical symmetry with
an assumed velocity field, and then iterated to match the
observed profiles. We discuss each of these procedures in turn.

\subsection{Static Chromosphere}

Two separate photospheric models were calculated with ATLAS
\citep{kurucz01}, one for stars in M13 and one for 
stars in M15 and M92. In order to create the initial photospheric
models, we used log~g=0.45, [Fe/H]=$-$2.45, 
T=4275K for the metal poor stars, and log~g=0.5, [Fe/H]=$-$1.5,
T=4500K for the metal rich stars. Although the
photospheric parameters of the target stars are different, 
this does not affect the calculated line profiles, 
because the H$\alpha$ line forms in the extended warm
chromosphere. The model atmospheres were represented by 72 
depths, where the photospheric distribution was given by the original Kurucz
values at the innermost 12 points. For the emission line calculations,
we changed the parameters at the outer depths to represent a
chromosphere with the temperature increasing linearly with decreasing 
mass column density [ZMASS (g cm$^{-2}$)]. Examples of these chromosphere 
can be seen in Figure 2 ({\it upper panels}) as a function of depth
index. In every model we assumed the stellar 
radius to be equal to 70R$_{\odot}$ and a microturbulent velocity of
4~km~s$^{-1}$ at each point of the 
atmosphere. We assumed that every star has a mass of 0.8~M$_{\odot}$,
making the gravity, $g$, also a constant parameter. 
This way the models depend on three free parameters: 1) the column
mass and temperature where the chromosphere starts; 2) the slope of the 
T$-$ZMASS function; 3) the highest temperature (T$_{\rm max}$) and the 
lowest ZMASS values of the chromosphere, 
where the transition region starts. For further simplicity, the column
mass and temperature where the chromosphere starts, and the mass column density 
where the chromosphere stops were also fixed. A transition region with 
a maximum temperature $2\times10^{5}$K was added to every model in 
the last 10 points to obtain small optical depths as hydrogen becomes
completely ionized. The chromosphere was represented at 
50 points linearly distributed in T vs. log ZMASS, which was
sufficient to sample in detail the region where the H$\alpha$ wings and core form. 

For every T$-$ZMASS distribution, we solved the non-LTE radiative
transfer and the statistical and hydrostatic
equilibrium equations, using the program PANDORA \citep{avrett01}. 
By keeping the starting T-ZMASS point, 
and the ending ZMASS point of the chromosphere the same, 
the only control parameter of the input models was 
the T$_{\rm max}$ value, which established the slope of the T$-$ZMASS
distribution. This parametrization allowed 
us to handle easily many different input models for the PANDORA program. 
These chromospheric models can be seen in Figure 3, and they are
listed in Table 2. We computed the 
non$-$LTE populations of a 15$-$level hydrogen atom assuming the same
value of the gravity, $g$, and [Fe/H] used in the photospheric 
models. All heavy elements were scaled using the metallicity used in the 
photospheric models and assuming the solar abundance distribution. 
For the continuum calculations we included all 15 bound-free transitions, and the most important
bound-bound transitions and scattering that contribute to the
photoionization of hydrogen. In every model we assumed a
microturbulent velocity of 4~km~s$^{-1}$ in each point of the atmosphere. 

Calculations were carried out in two phases for all models: in the
first phase a plane-parallel approximation was used in order to calculate the scale
of the atmosphere and the total H density from the initial and fixed
ZMASS values, the Kurucz scale height, and total H density. A run was
considered converged if the new height scale and total H density did 
not change by more than 1$\%$ as compared to
the previous run. After this, the plane-parallel atmosphere
was replaced with a spherical atmosphere with the
same stratification, and this spherical model was used to 
calculate the emergent spectrum. To check the 
accuracy of our input approximations, we changed the input radius, 
gravity, and [Fe/H] each by a factor of
2 for one model. Changing the [Fe/H] does not affect the line
profiles, but the radius and gravity do. A 
smaller radius (and larger gravity) leads to stronger emission. 
Although the line profile changes, the changes in radius and 
gravity do not affect the derived mass loss 
rates by more than our errors from the velocity-field determination (see next section). 

\subsection{Expanding Atmosphere}

From Figure 3, one can see that emission wings arise in warm, static chromospheres. 
However, the static atmosphere cannot explain the asymmetry of the
emission wings and the `banana-shaped' bisector of the observed
profiles \citep{dupree01}. Thus, flow velocities must be present in the atmosphere. 
Accordingly, the regions where the core and wings of the H$\alpha$
line are formed were put in motion. 
We constructed velocity distributions in order to produce asymmetrical 
line profiles to match the observed line asymmetries. This velocity
field is included when calculating the line source function. 
The line-forming regions were determined from the depths where the
maximum contributions to the spectrum occur. The H$\alpha$ line core 
forms between depths 16$-$21 (8000$-$9900 K), the wings from between depths 24$-$35 (5800$-$7800 K) 
in every model, depending on T$_{\rm max}$. The rest of the atmosphere does not affect the line profile, 
thus we do not have information on the velocity field 
outside these regions. The velocity was changed usually between $-$14 and 20~km~s$^{-1}$, where the
negative number means an inward velocity and the positive number means
an outward velocity relative to the photosphere. 

In order to match the line profiles, three characteristics of
H$\alpha$ were considered: 1) the
bisector velocity ($v_{\rm bis}$) and the position of the H$\alpha$ core, 2) the
width of the H$\alpha$ absorption line,  
and 3) the ratio of of the strength of the blue and red emission wings (B/R). The velocities of 
$v_{\rm bis}$  are calculated in the following way
\citep{meszaros02}: the absorption line is divided into about 20
sectors of equivalent depth; the top and 
the lower 3 sectors are selected and the wavelength average of the top
and lower sectors is calculated, subtracted 
one from another and changed to a velocity scale.
The fitting was done by, first, taking a well converged spherical run in which the modeled 
emission matched the observation. The strength of the emission has an important 
effect on the calculated mass loss rate, because in our approximations it scales the atmosphere. 
Higher T$_{\rm max}$ corresponds to a larger height scale (Figure 2, lower panels), and smaller hydrogen density, 
because the mass column density is the same in every model. 
Then, the previously measured bisector velocity of the
H$\alpha$ absorption line gave an estimate of the expanding velocity in the core. However, in almost every case
this velocity did not yield the same bisector and position of the core in the calculated profile as in 
the observed one. Higher values of the velocities were required
indicating that the measured H$\alpha$ bisectors are a lower limit to the actual velocity fields present in
the star. Nearly a factor of 2 higher velocities were necessary in the
models in every case. The value of 
velocity in the region where the wings form influences the B/R ratio. In most cases, if B/R$>$1, then an 
inward velocity was needed; when B/R$<$1, an outflowing velocity was required. Our observations show that the
H$\alpha$ line cores are either at rest or moving outward from the star. Moreover, the line asymmetries can
change from one observation to another, thus a complex time-dependent velocity field must be present in these
chromospheres. 
				   
The microturbulent velocity was also changed from 4~km~s$^{-1}$ 
(used in the generic spherical calculations) in order to match the width along the H$\alpha$ line with the 
observation. The assigned microturbulent velocity varied between 6 and 14~km~s$^{-1}$ in the region where 
the wings of the  H$\alpha$ line form, and 0~km~s$^{-1}$ where the core
forms in order to better match the width of the core.  
After the calculation was completed, the model profile was compared with the observation by eye and further 
adjustments were made to the velocity field. This was continued until the modeled and observed line profiles 
matched each other as well as possible. The mass loss was then estimated with a simple formula based on 
mass outflow:
\begin{center}
$\dot{M}$(M$_{\odot}$~yr$^{-1}$) = 2.33 $\times$ 10$^{-26}$ $\times$ m$_H \times$ N$_H
  \times$ 4 $\pi \times$ R$^{2} \times$ V$_{exp}$
\end{center}
where $m_{H}$(g) is the mass of the 
hydrogen atom, $N_{H}$(cm$^{-3}$) is the total hydrogen density, $R$(cm) is the distance from stellar center, 
and $V_{\rm exp}$(cm~s$^{-1}$) is the velocity of the outermost layer.  
In our calculation the velocity, distance, and hydrogen density of the
outermost layer forming the line core gave the mass loss rate for each star. 

To obtain an estimate of the error of the mass loss rates, the expanding velocity in the wing and core forming 
regions was changed by $\pm$1~km~s$^{-1}$ in every depth. In both cases the mass loss changes by 
less than a factor of 2 and the line profile changes are not visible by eye. Changing every depth by 
$\pm$2~km~s$^{-1}$ usually gave a worse fit to the observation; thus we conclude that the error in the derived
mass loss rates appears to be a factor of 2. 

\section{Discussion}

\subsection{Observed and Calculated Profiles}

The comparison of observed and calculated spectra is shown in Figures
4$-$9, and the derived mass loss rates are 
listed in Table 2. The spectra were computed at high resolution, and were convoluted with a Gaussian 
distribution corresponding to the spectral resolution of $\sim$34000. We aim to match several of the line 
parameters: the central core depth, the core velocity shift, the line width, the strength and
the asymmetry of the emission wings. Changing the model usually produces changes in more than one line 
parameter, and so our final model, the `best fit', is frequently a compromise solution. 
In some cases the continuum level of the observed spectrum was shifted to match the 
calculated one, and we were able to match the observed profiles fairly well. 
The main difference between calculations and observations is that the computed H$\alpha$ profiles are
slightly broader and deeper in the core; similar systematic differences were found by 
\citet{mauas01}. This suggests that in our models there is either
slightly more hydrogen in the atmosphere where the H$\alpha$ core forms, or the chromosphere is hotter, thus
increasing scattering from the core. The calculated atmospheres are 
homogeneous in every case, so that the difference in the core 
might also come from  inhomogeneities in the atmosphere.  
This, however, does not affect the mass loss rate calculations by more than a factor of 2, because the wing 
asymmetry, bisector, and the position of the core are taken into account in fitting the observed profiles, and
these characteristics are more important in determining the calculated mass loss. This 
homogeneous chromosphere approximation gives better results for brighter objects (for example L973 in M13, 
Figure 4) and only the faintest stars in each cluster (L592 in M13, K87 in M15 and XII-34 in M92) show major 
differences in the H$\alpha$ core. In some cases an inward velocity
had to be used where the emission wings formed in order to 
match the wing asymmetry, but in all cases an outward velocity was necessary to fit the core. 

The stars, L592, K87, and XII-34 did not show any emission in H$\alpha$ and have lower luminosities than other 
selected RGB stars with H$\alpha$ emission. This makes it difficult to derive a mass loss rate for these 
stars. It was not possible to construct an accurate 
model when emission is not visible in the spectrum, because in our approach the emission wings were used to 
give the slope of the temperature versus mass column density in the static chromosphere. 
Thus only the bisector and the core of the 
H$\alpha$ line affect the fit. The difference between the observed and calculated profile also derives from 
the fact that the radius and surface gravity for these stars are quite different (much smaller) than we
assumed in each calculation. In order to check the accuracy of the derived mass loss rate, a 
static, spherical chromosphere was calculated using R=35R$_{\odot}$ and log~g=1.25 $--$  parameters 
close to the values for these stars. The same velocity field was applied to this chromosphere as to the others. 
The mass loss rates do not differ from each other by more than a factor of 2, but the new profiles using the
smaller radius do not match the observations very well. Thus, in the final interpretation here we use the model 
with R=70R$_{\odot}$. 

The H$\alpha$ core generally forms between T=8000 and 9900K, which, in our models, is located between 1.4 
and 2.0~R$_{*}$ in the chromosphere. All of the observed H$\alpha$ profiles have a static or outflowing 
core. The semi-empirical models thus all require outflow at the top of the atmosphere in order to match 
the profiles. At the highest temperatures ($T > 10^4$K), there is no 
contribution to the H$\alpha$ profile, hence we have 
no information on any velocity field that might be present. Thus, we have reduced the velocity to zero.
At lower levels in the atmosphere, below T$\sim$8000K either inflow or outflow occurs. 
The direction of the velocity field in the model is determined by the
asymmetry of the line wing emission. When B$<$R, an outflowing velocity is required; when B$>$R, an 
inflowing velocity is required.

Outflowing velocities used in the modeling vary between 3 and 
19~km~s$^{-1}$, which are much
smaller than the escape velocity (50$-$70~km~s$^{-1}$) from this part of the chromosphere (Table 2). 
In our spectra, the H$\alpha$ core is either at zero velocity with respect to the star, or moving outwards. 
There are no signs of any inflow in the core itself. Therefore it appears reasonable to assume that 
the outward velocity continues to increase until the escape velocity is reached. While the 
material is not escaping from the chromosphere where the H$\alpha$
core forms, analysis of \ion{Mg}{2} lines and 
the \ion{He}{1}~$\lambda$10830 absorption line of RGB stars in NGC~6752, 
M13, and metal-poor field giants shows that velocities can reach up to
140~km~s$^{-1}$ \citep{dupree1992, dupree02, dupree2007, dupree2009, smith03}. These lines 
are formed higher in the atmosphere than H$\alpha$, which suggests that the mechanisms driving the 
stellar winds become stronger above the top of the chromosphere and escape of material is only 
possible at distances $>$2.0~R$_{*}$.

\subsection{Changes in Time}

One star in M13 (L72) and M92 (VII-18) and three stars in M15 (K341, K757, and K969) were observed more than
once, which allows us to examine how the mass loss might change between observations. Separate semi-empirical
atmospheric models were constructed to match each of the observed profiles. 
Of these 5 stars, three showed evidence of a difference in the mass
loss rate, and two had nearly the same mass loss rate between  observations.  
In the case of K341 (Figure 6) the mass loss rate changed only slightly with the respect to the observed 
spectrum, and these differences are smaller than the error of the mass loss determination. L72 showed less 
than a factor of 2 change in the mass loss rate over a month time span
even though the observed spectra are quite different in the asymmetry of the emission wings
(Figure 4). Although the atmospheric motions changed, this created only a slight
difference in the mass loss rate. A much larger difference occurred in K757 (Figure 7), where the core asymmetry
became prominent, and the derived mass loss rate increased from 5.7$\times$10$^{-10}$ 
M$_{\odot}$~yr$^{-1}$ to 3.0$\times$10$^{-9}$ M$_{\odot}$~yr$^{-1}$ -- by almost a factor of 6. 
Nearly one and a half years separate these two observations. Such large changes were visible in AGB stars in M15 
\citep{meszaros02} near log~$(L/L_{\odot})\sim2.0-2.7$ in 1.5 years. This star, K757, is however on 
the RGB according to its position on the CMD (Figure 1) and was not identified as a dusty AGB star by 
\citet{boyer02}, which demonstrates that large changes in the mass loss rate can occur on the RGB as well. 

Of these 5 stars, two showed evidence of pulsation in the spectra. The B/R ratio of 
emission in the spectra of K341 changed between observations. During
the first observation (2005 May) the emission asymmetry signals 
inflow; in the following 3 observations (2006 May and October) the
asymmetry suggests different values of outflow. Although in the case 
of the first observation the modeled velocity is slightly greater 
than zero, the emission ratio cannot be modeled with 
outflow velocities. The H$\alpha$ core, however, shows an outflow in every observation. We take this as 
evidence of a pulsation present in the lower chromosphere. Our models show that the pulsation
extends outwards into the chromosphere to around 1.4$-$1.5~R$_{*}$; at these
levels and below, both inward and outward flows 
are possible, but the higher parts of the chromosphere ($>$1.5 R$_*$) do
not participate in the pulsation. It is more likely that 
pulsation itself helps to drive the mass outflow. 
The only other star with varying inflow and outflow velocities where
the emission forms is K969. The dust-free models of \citet{struck01}
suggest that a stellar wind may be supported by shock waves which travel through the wind, possibly
related to the pulsation in the lower levels of the atmosphere, similar to what is present here in the red giants.

\subsection{Dependence on Luminosity, Effective Temperature and Metallicity}

Although the bisector velocity was one of the key parameters taken
into account while fitting the models to 
the observations, the relation between the bisector velocity and the derived mass loss rate is not 
unambiguous. Generally the semi-empirical models require higher
expansion velocities than the measured bisector velocity. 
In some cases, matching a  spectrum exhibiting a low bisector velocity ($<-$2~km~s$^{-1}$) 
is only possible with models assuming high outflow velocities ($>$10~km~s$^{-1}$) in the core.
This usually occurs when either the blue or the red emission is weak. Also, 
while calculating the bisectors of the line profiles, the top 
sector was close to the continuum in the normalized spectrum, but the lowest sector was placed 0.01~$-$~0.05 
above the lowest value of the line depending on the signal-to-noise ratio in the line core. In these
cases, matching the position of the core was significant in the fitting, which was only possible  
using high outflowing velocities ($>$10~km~s$^{-1}$).

The mass loss rate weakly increases with luminosity and with decreasing effective temperature in 
M15 and M92 (Figure 10). This is because the bisector velocity increases with luminosity (decreases
with effective temperature), and while the atmospheres of these stars
are larger and thus less dense, the increase in 
expansion velocity necessary to reproduce the absorption line profile gives an overall higher mass outflow. 
This weak dependence is affected by the error of modeling and 
the changes with time in the dynamics of the atmosphere. It is also evident that stars in the more metal rich 
cluster, M13, have larger mass loss rates than in the metal poor clusters M15 and M92, although this 
difference is close to our errors. This is expected from the previously observed bisector 
velocity$-$luminosity relation \citep{meszaros02, meszaros01}, because the H$\alpha$ line bisector 
velocity was one of the key fitting parameters. In M13, low (close to
$-$2~km~s$^{-1}$) bisector velocities were measured for 
the most luminous stars (L954, L973, log~$(L/L_{\odot})>3.3$, Figure 4) from H$\alpha$ and because bisector 
velocity was a significant fitting parameter, the derived mass loss
rates are comparable to those for the fainter 
stars. Although the bisector velocities of the H$\alpha$ core approach
$-$2~km~s$^{-1}$ as compared to $-$6 to $-$9~km~s$^{-1}$ for stars 
between log~$(L/L_{\odot})=3.0$ and 3.3, the mass loss rate did not
decrease significantly. Our spectra showed \citep{meszaros01} that the 
velocities as determined from the H$\alpha$ profile decrease at the highest luminosity which
may result from the changing atmospheric structure in the most
luminous stars. It does not seem physically reasonable that the
outflow would cease at the highest luminosities. The mass loss rate did not
change greatly for these objects, because relatively high velocities
are needed to use to match the position of the core 
in the H$\alpha$ line. In sum, the values of the rates that we find
are generally similar to the non-LTE studies of H$\alpha$ in other metal-poor giants
\citep{dupree01,mauas01}.

\subsection{Comparison with Other Models and Mass Loss Relations}

The values of the mass loss rates that we find range from 
0.6$\times$10$^{-9}$ to 5 $\times$10$^{-9}$ (M$_\odot$~yr$^{-1}$).
These values are in general agreement with other calculations for 
metal-deficient giant stars in the field or in the cluster NGC 2808.
\citet{mauas01} modeled five very cool stars (T$_{eff} <$4015K) and
found differences in the mass loss rates amounting to a factor of 38
-- with values ranging from 0.1$\times$10$^{-9}$ to
3.8$\times$10$^{-9}$ M$_\odot$~yr$^{-1}$. 
This may reflect the sort of episodic change that
might lead to dust production. However, because the line cores
of H$\alpha$ in the models do not seem to well match the
observed profiles, the derived mass loss rates could be affected.
Their velocity profiles also differ from ours. In fact, 
inward velocities were not required to match the 
observed H$\alpha$ emission line profiles, even if they signaled inflow. 
For line profiles with B$>$R, they introduced a
decelerating velocity field with increasing radius in order to match
the profiles. Thus, the zero point of the chromospheric velocity is different.

The literature contains various relationships to estimate
the mass loss rates for luminous cool stars. The widely used
`Reimers law' based on dimensional arguments \citep{reimers01,reimers02}, 
derives from a handful of Population I giant stars. This mass loss formula 
was later revisited by \citet{catelan01} and suggested a stronger dependence on luminosity, radius and log g. 
The SC relationship \citep{schroder01} is more detailed and includes gravity and effective temperature and
assumes that the wind arises from an extended, highly turbulent chromosphere, possibly associated with Alfven 
waves. They did not consider extremely metal-deficient stars, such as
those in M15, in their calibration. \citet{origlia02} presented an 
empirical (dusty) mass loss formula based the globular 
cluster 47 Tuc. Dusty RGB stars were identified from  
mid-IR photometry with \textit{Spitzer}. Mass loss rates were 
calculated by modeling the emerging spectrum and dust emission with the DUSTY code 
\citep{ivezic01, elitzur01}. The dependence of the mass loss rate on luminosity is much shallower than 
suggested by the Reimers relationship.

These rates are shown in Figure 10 and listed in Table 3, where they differ from our
model calculations by an order of magnitude at least. For the faintest stars, below log~$(L/L_{\odot})=2.8$ 
the Schr\"oder-Cuntz (SC) relation predicts the lowest values of the 3 approximations;
at higher luminosities the difference between the SC relation and our model calculations increases, amounting to an
order of magnitude at the highest luminosities. 
While the slope of the Origlia relationship with luminosity is similar to
ours, the predicted values of the mass loss rate are larger by 
more than an order of magnitude. Their formulation included a scaling
factor, $C$, which is the product of the 
gas-to-dust ratio, the expansion velocity, and the grain density. They set
$C=1$ for 47 Tuc. Our clusters are lower in metallicity, presumably increasing
the gas:dust ratio and the expansion velocities are slightly higher than
the 10 km s$^{-1}$ taken by \citet{origlia02}. Thus, for the same
grain density, the discrepancy between the dust rates and the modeled
H$\alpha$ rates would increase. 

The rates derived for dusty winds from mid-IR photometry are 
consistently higher than those indicated by the gas. If the dust is
produced episodically \citep{meszaros02,origlia02} at these high
rates, it is puzzling that anomalously massive outflows have not been detected in
the optical spectra. The IR observations led \citet{origlia02} to
conclude that mass loss in 47 Tuc is ongoing in a fraction of the stars ranging
from 16 to 32 percent on the RGB, whereas the H$\alpha$ spectra and
modeling of stars in M13, M15, and M92 shows that all of the stars have outflowing chromospheric
material. 

Models of two stars in M15, K421 and K479, previously identified as dusty AGB stars by \citet{boyer02} were
also calculated (Figure 9). These stars are similar in luminosity, effective temperature, and bisector 
velocity to other RGB stars included. For these two stars we find no difference in mass loss rate 
from other red giants. Mass loss rates suggested by the IR excess exceed by more than an order
of magnitude the rates inferred from H-alpha. If the mass loss must be
high in order to produce dust, we conclude that the M15 giants are not 
currently undergoing an episode of dust-production. The dust observed in the dusty RGB stars most likely 
left the star decades earlier so one does not necessarily expect a correlation
between time varying chromospheric phenomena and dusty envelopes. 

A least-squares fit to our mass loss rates as a function of
luminosity, temperature and [Fe/H] yields the following form:
\begin{center}
$\dot M$[M$_{\odot}$~yr$^{-1}$] = 0.092 $\times$
L$^{0.16}\times$ T$_{\rm eff}^{-2.02}\times $A$^{0.37}$ where A=10$^{[Fe/H]}$.
\end{center}
Here we have excluded the 2 most luminous stars of M13 from the fit,
because we believe that H$\alpha$ becomes less sensitive to the mass
outflow at the low temperatures of the metal-rich red giants. The
values calculated from this relationship are given in Table 3 and shown in Figure 10.

\section{Conclusions}

1. Chromospheric modeling of the H-alpha line in several clusters
demonstrates that the mass loss rate increases with increasing luminosity and
decreasing effective temperature of stars on the red giant branch. 
All stars modeled down to 2 magnitudes
below the RGB tip show outflowing material suggesting that mass loss is a continuous process. 
The more metal-rich stars have a higher mass loss rate than the metal-poor stars.
We offer a new relationship for mass loss rates in Pop II stars based
on these models.

2. The calculated mass loss rates from the H$\alpha$ profile give values that are an order of magnitude less
than those estimated from the \citet{reimers01,reimers02}, SC
\citep{schroder01}, and \citet{origlia02} relationships. Differences are larger at 
higher luminosities. The H$\alpha$ mass loss rates and the Origlia relationship give a very similar shallow
dependence on luminosity. 

3. At the top of the RGB, for stars brighter than log~$(L/L_{\odot})=3.3$, the H$\alpha$ line may not be 
adequately sensitive to the mass loss rate; the models suggest lower mass loss rates for these
objects.

4. An expanding velocity at the top of the atmosphere was required for
every star in order to match the 
H$\alpha$ core. The largest outflowing velocity reached 19~km~s$^{-1}$, usually larger by factors up to 
10 than indicated by the bisector velocity. 
In the region where the H$\alpha$ emission is formed, the velocities
can change direction, indicating the presence of pulsation. 

5. Two stars previously identified as dusty red giant stars in M15 show no difference in mass loss rate 
from other red giants. If high rates of mass loss are needed  
in order to produce dust, we conclude that the M15 giants are not 
currently undergoing an episode of dust-production.  

6. K757 (M15) shows a factor of 6 mass loss$-$rate change in a
time span of 18 months (from 
5.7$\times$10$^{-10}$~M$_{\odot}$~yr$^{-1}$ to
3.0$\times$10$^{-9}$~M$_{\odot}$~yr$^{-1}$). A smaller change occurred in 
two other stars, K341 (M15) and L72 (M13), where the mass loss
difference was nearly a factor of 2. 

\acknowledgments{We are grateful to Rudi Loeser who was instrumental
in developing the PANDORA code. We thank Bob Kurucz for help with the 
model photospheres.  Szabolcs~Meszaros is supported in part 
by a SAO Predoctoral Fellowship, NASA, and the Hungarian OTKA Grant 
K76816. This research is also supported in part by the Smithsonian Astrophysical Observatory.}

\clearpage

\begin{figure}
\includegraphics[width=4.5in,angle=270]{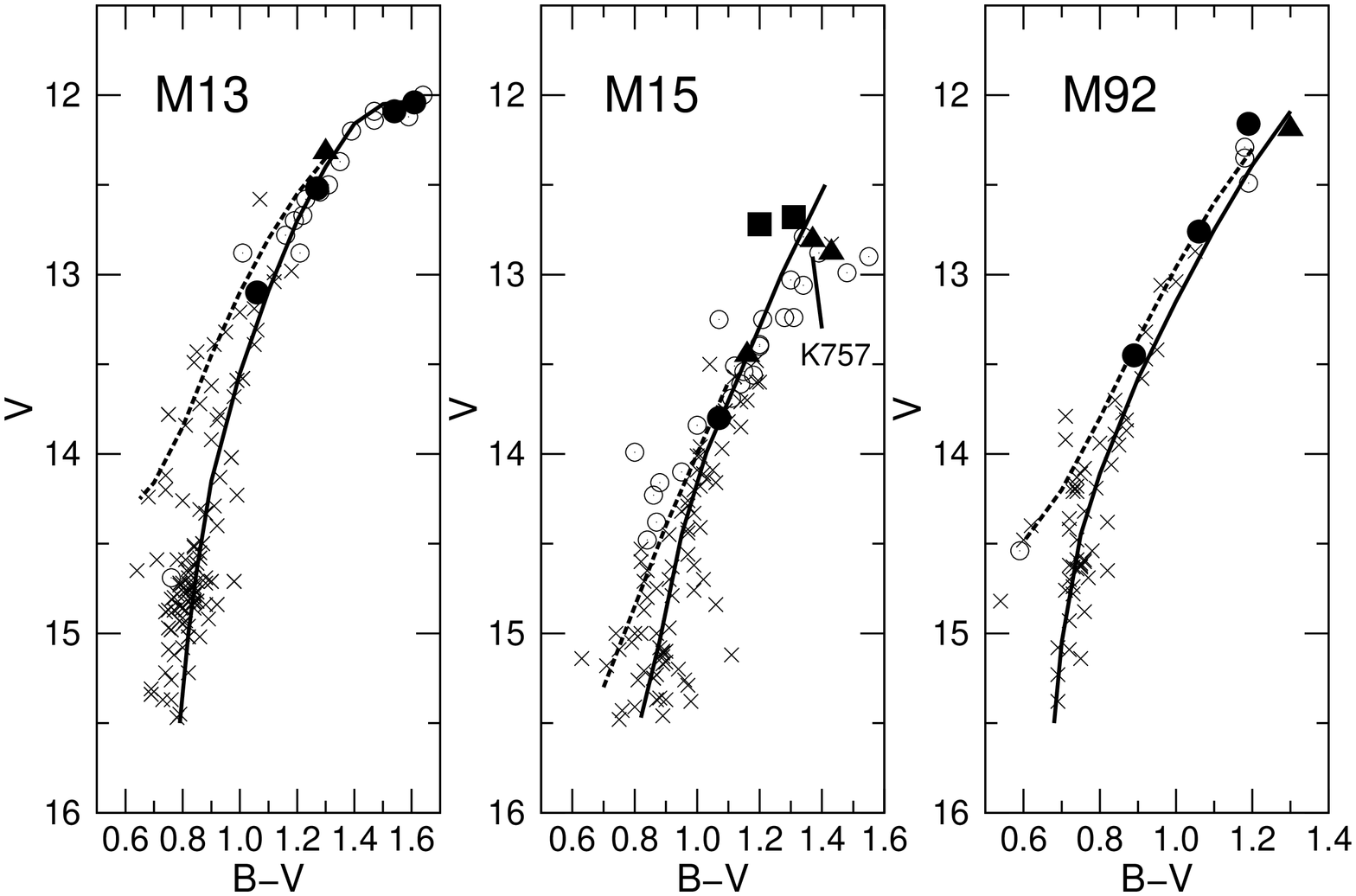}
\caption{Color-magnitude diagram for all stars observed in M13, M15 and M92. Stars with H$\alpha$ emission are marked
by open circles. Filled symbols mark the stars modeled in this paper, where stars observed once are marked by filled
circles, stars observed more than once are marked by filled triangles, and stars observed with \textit{Spitzer} 
\citep{boyer02} are marked by filled squares. 
The solid line shows the fiducial curve of the RGB; the dashed line traces the fiducial curve of the 
AGB for M13 and M92 taken from observations of \citet{sandage01}, and for M15 
taken from observations of \citet{durrell01}.
}
\end{figure}

\clearpage

\begin{figure}
\includegraphics[width=4.5in,angle=270]{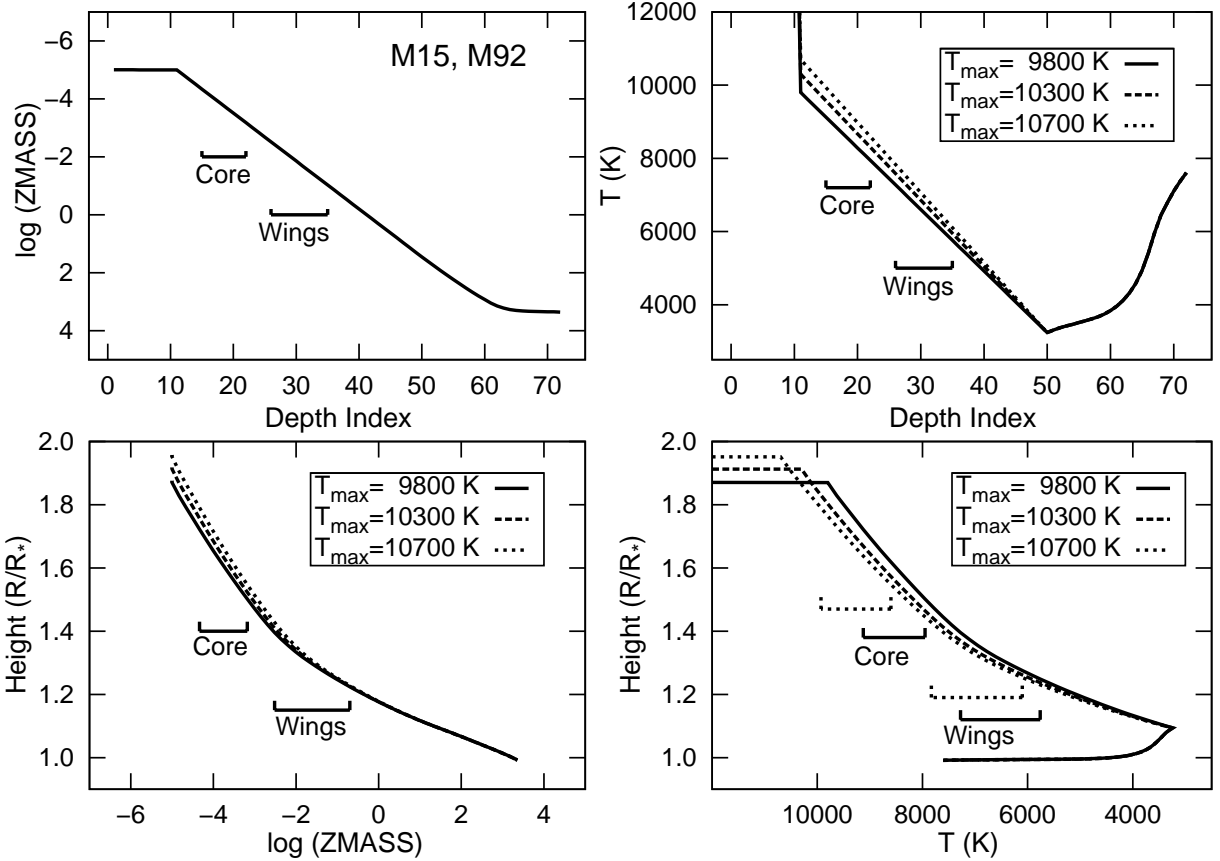}
\caption{{\it Top left and right:} The mass column density and temperature of three selected input models as a 
function of depth index. The atmosphere was sampled with 72 points, the mass column density was kept the 
same in every input model. The depth index equals 0 at the top of the chromosphere and increases downward
through the chromosphere and the photosphere. The line formation regions were determined from the maximum
values of the contribution to the line profile. 
{\it Lower left and right:} The height of the chromosphere as a function of  mass column density and temperature.
The height was calculated assuming a R=70R$_{\odot}$ radius. The H$\alpha$ core forms between depths 16-21 
(8000$-$9900 K, depending on T$_{\rm max}$); the wings form between depths 24-35 (5800$-$7800~K, depending on 
T$_{\rm max}$). See Section 3 for more information. 
}
\end{figure}

\clearpage

\begin{figure}
\includegraphics[width=4.5in,angle=270]{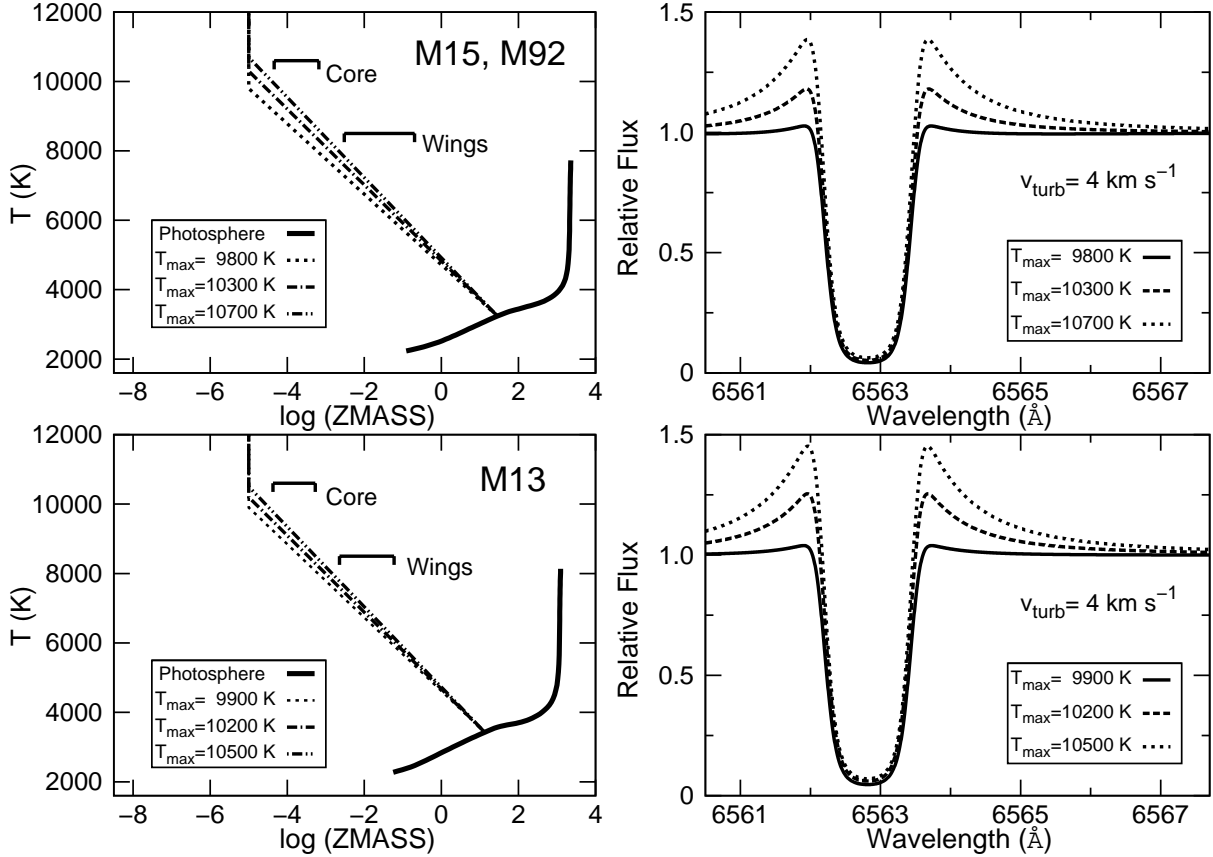}
\caption{{\it Top left:} The temperature distribution of the chromosphere as a function of mass column density 
for three of the input models for M15 and M92. The solid line is the Kurucz model without any chromosphere. 
In all cases the chromosphere extends to ZMASS=$1\times10^{-5}$ (g cm$^{-2}$), where the transition 
region starts. T$_{\rm max}$ is the maximum temperature of the chromosphere. The regions of
formation for the  H$\alpha$ wing and core are marked. 
{\it Top right:} H$\alpha$ profiles for 3 models.
Only a few hundred K differences in the maximum temperature
result in large changes in the emission. The three static input
models use 4~km~s$^{-1}$ for the turbulent velocity, constant with depth. 
{\it Lower panels:} The same results for the models of stars in M13. 
}
\end{figure}

\clearpage

\begin{figure}
\includegraphics[width=5in,angle=0]{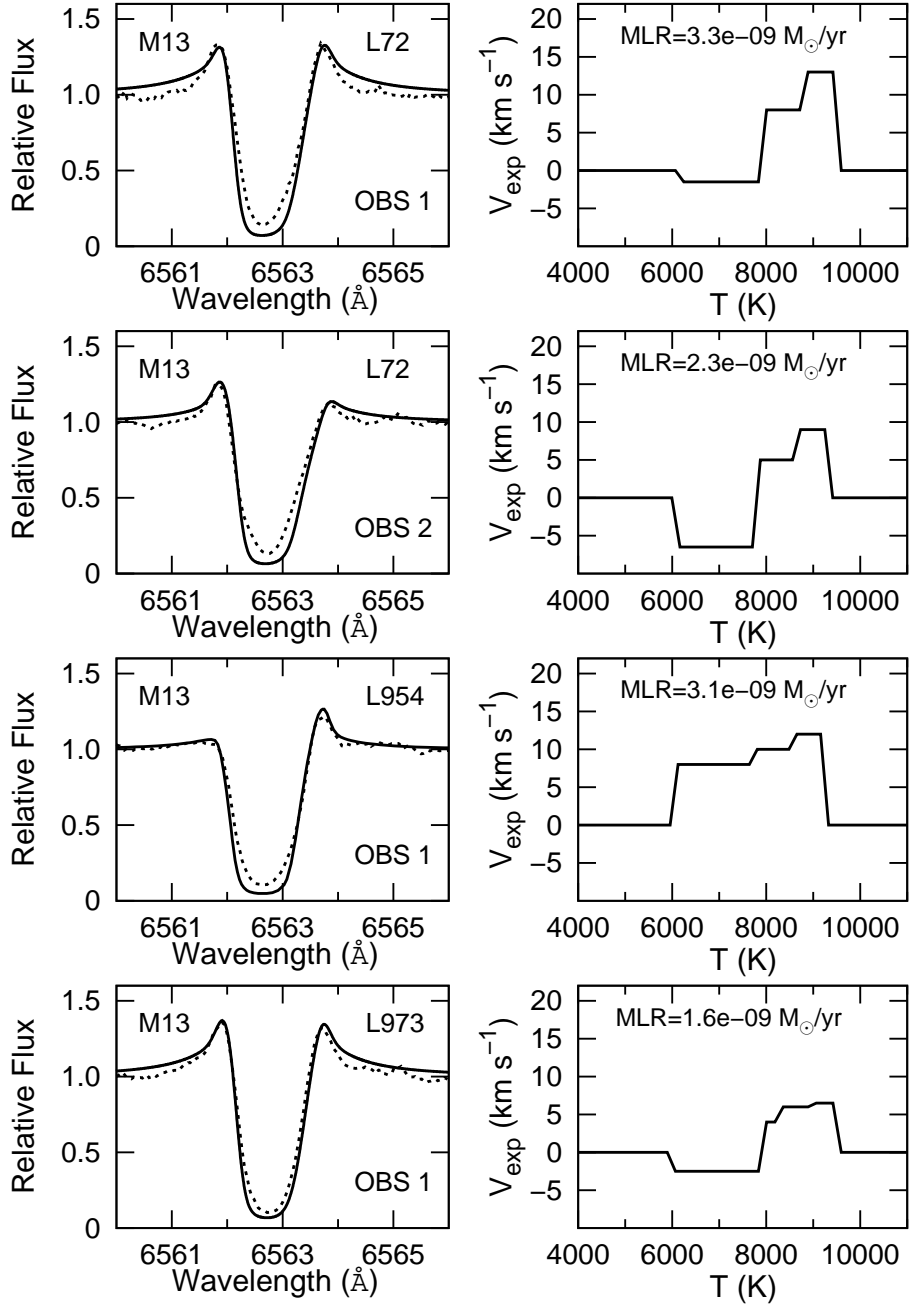}
\caption{{\it Left panels}: Calculated spectra compared to the
  observations of stars in M13. The solid line shows the calculated line
profile; the broken line marks the observation. {\it Right panels:} The expansion velocity (v$_{exp}$) used to
match the line profile as a function of  temperature. The expansion velocity is positive
for a outwardly moving flow and  negative for a inwardly moving flow.
  The derived mass loss rate is indicated for each model. 
}
\end{figure}

\clearpage

\begin{figure}
\includegraphics[width=5in,angle=0]{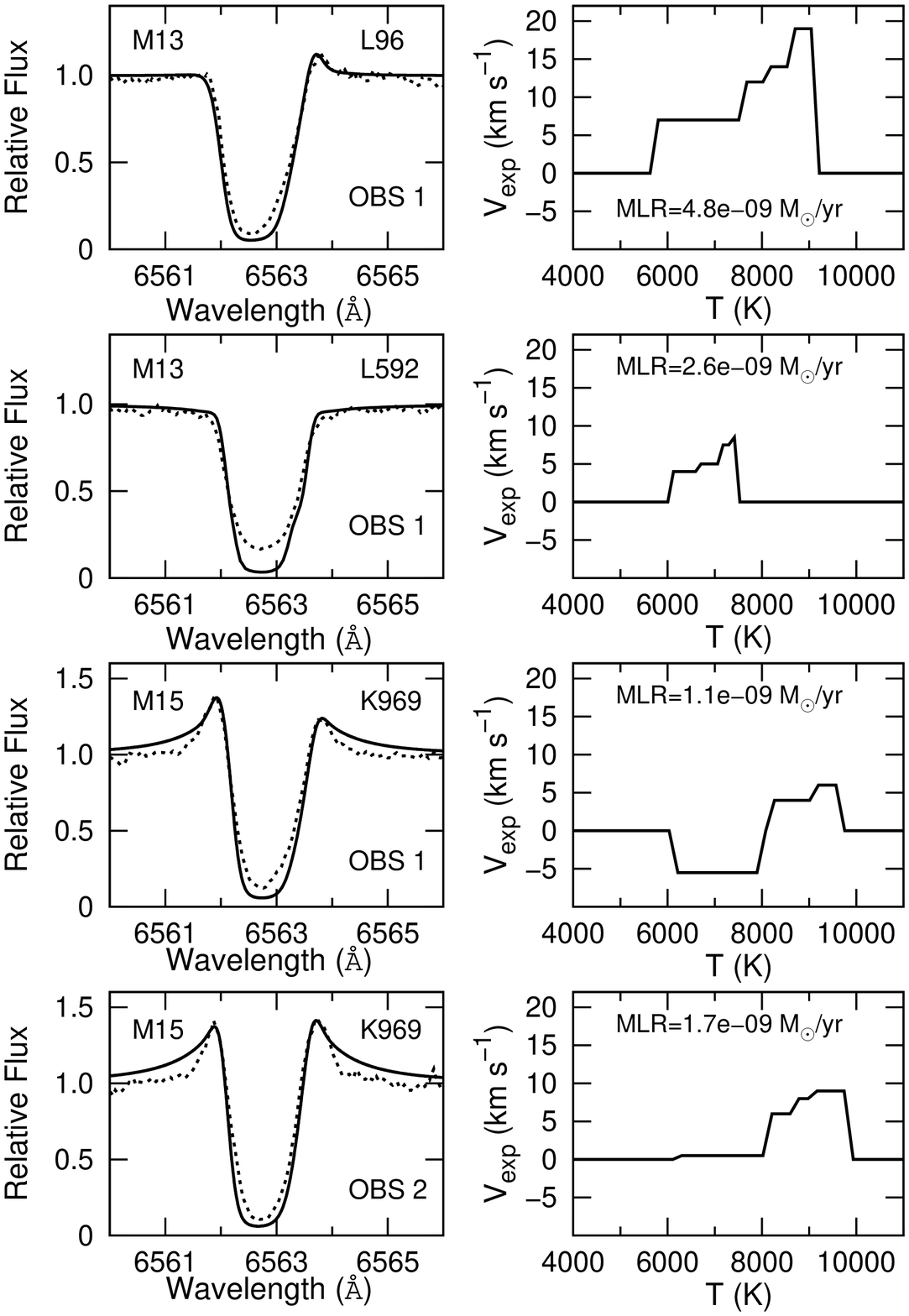}
\caption{{\it Left panels:} Calculated spectra compared to the
  observations. {\it Right panels:} The expansion velocity used to
match the line profile as a function of  temperature. For additional explanation see the
caption of Figure 4. 
}
\end{figure}

\clearpage

\begin{figure}
\includegraphics[width=5in,angle=0]{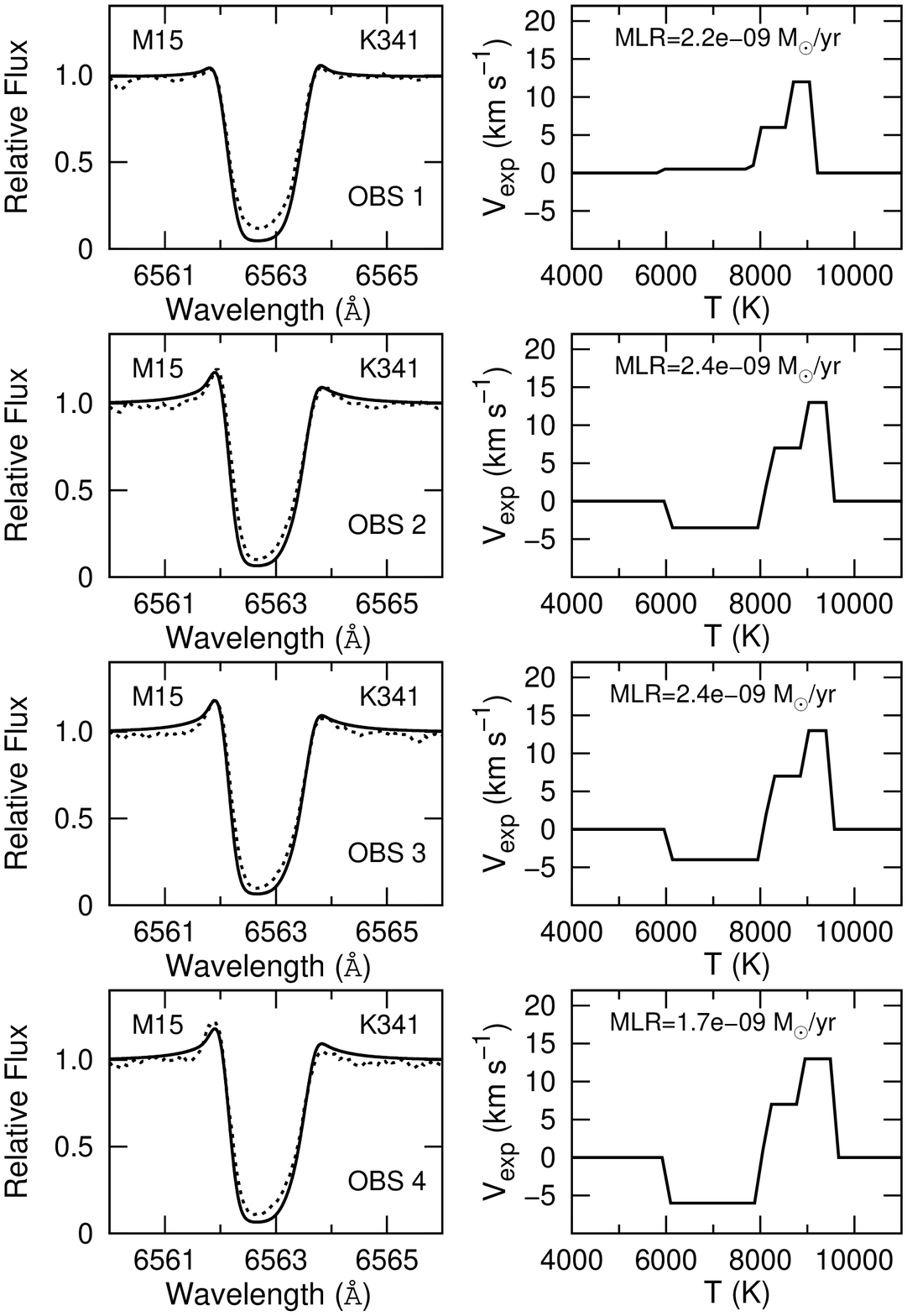}
\caption{{\it Left panels:} Calculated spectra compared to 
the observations. {\it Right panels:} Expansion velocity used to
match the line profile as a function of the temperature of the chromosphere. For additional explanation see the
caption of Figure 4. 
}
\end{figure}

\clearpage

\begin{figure}
\includegraphics[width=5in,angle=0]{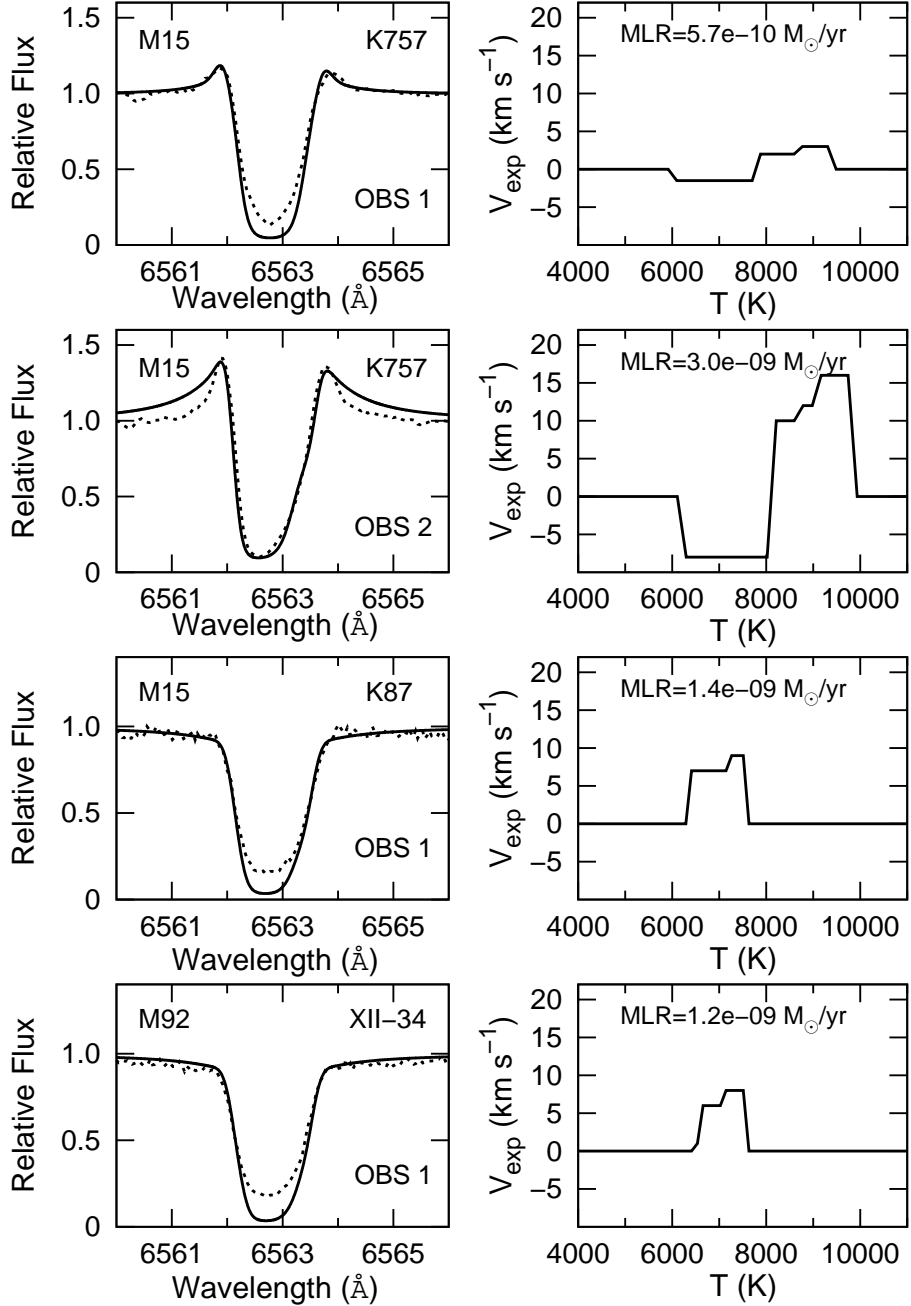}
\caption{{\it Left panels:} Calculated spectra compared to the
  observations. Note the difference in the profiles between the two
  observations of K757. {\it Right panels:} Expansion velocity used to
match the line profile as a function of the temperature. For additional explanation see the
caption of Figure 4. 
}
\end{figure}

\clearpage

\begin{figure}
\includegraphics[width=5in,angle=0]{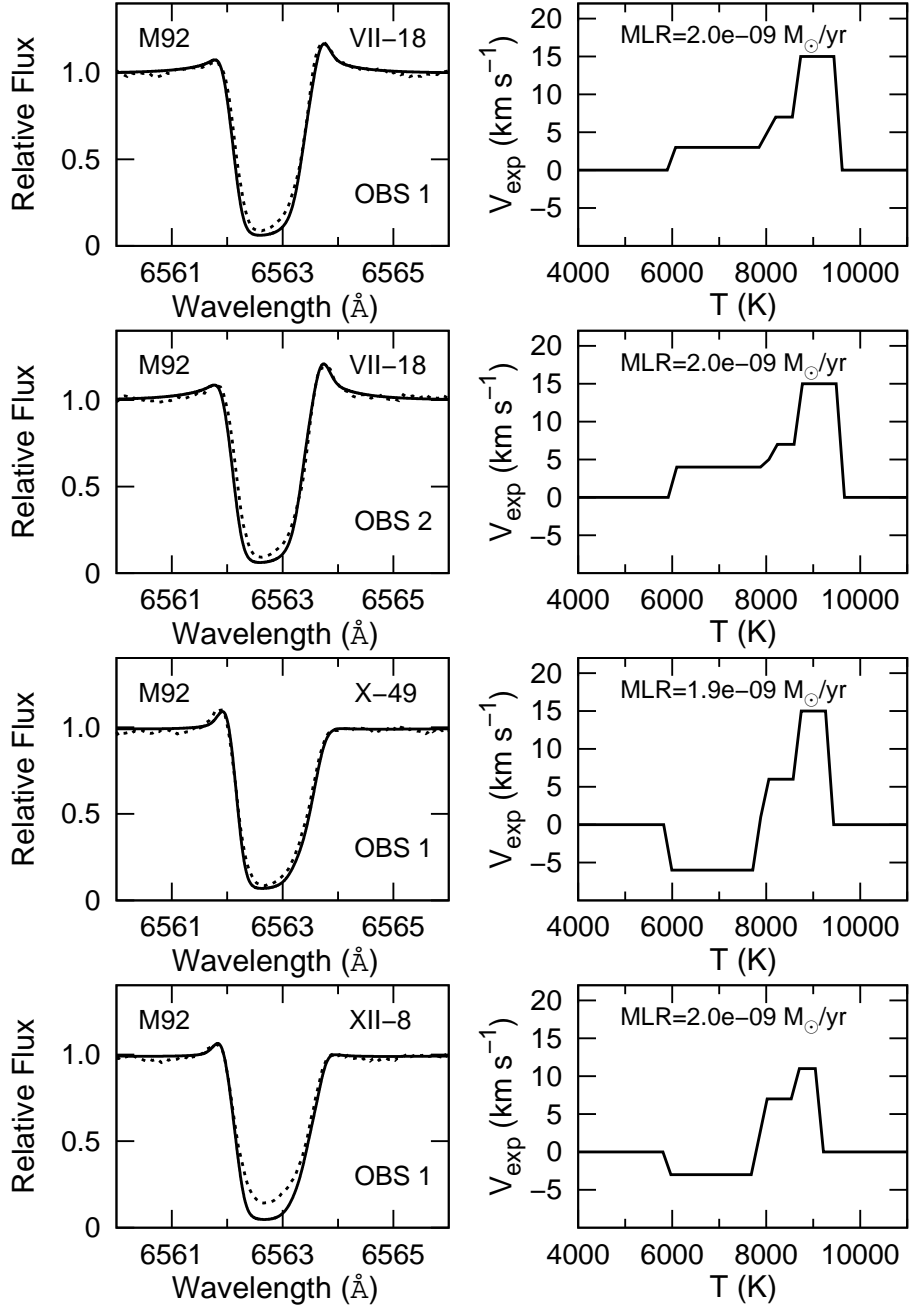}
\caption{{\it Left panels:} Calculated spectra compared to the
  observations. {\it Right panels:} Expansion velocity used to
match the line profile as a function of the temperature. Additional 
explanation can be found in the caption of Figure 4. 
}
\end{figure}

\clearpage

\begin{figure}
\includegraphics[width=4.5in,angle=270]{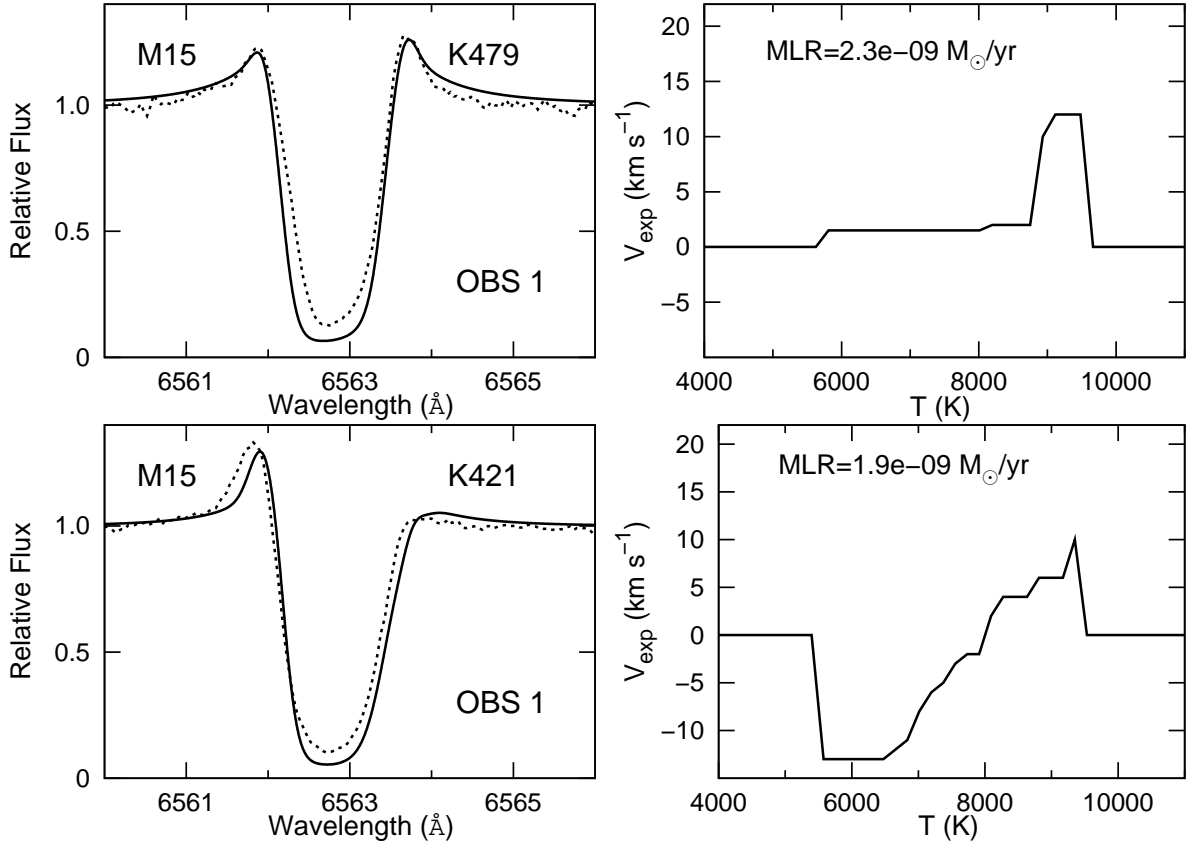}
\caption{{\it Left panels}: Calculated spectra of giant stars identified as having circumstellar material from
\textit{Spitzer} observations \citep{boyer01} compared to the observations. 
{\it Right panels:} The expansion velocity used to match the line
profile as a function of temperature. Additional explanation can be
found in the  caption to Figure 4. 
}
\end{figure}

\clearpage

\begin{figure}
\includegraphics[width=4.5in,angle=270]{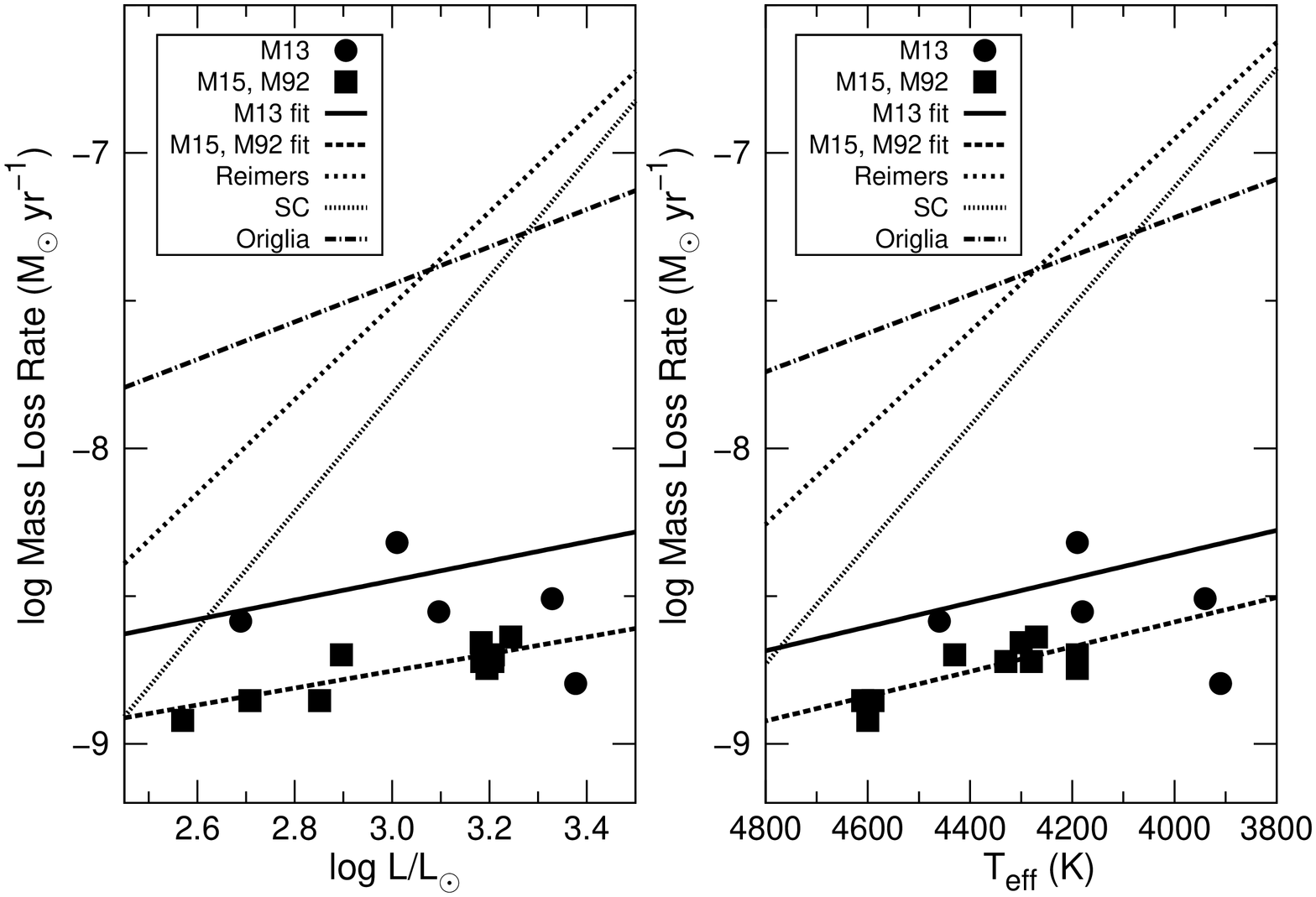}
\caption{Average mass loss rates calculated in this paper ({\it solid circles} 
and {\it solid squares}) as compared to relations 
proposed by \citet{reimers01, reimers02},  
\citet{schroder01}, denoted by SC, and \citet{origlia02}. The C
parameter introduced by \citet{origlia02} was set equal to 1. Mass loss rates 
for each of the target stars were calculated from the 3 relationships, and the 
curves shown were fit to the individual points. Our derived mass loss rates 
from the H$\alpha$ line profiles are almost a factor of 
10 smaller than from existing mass loss approximations. The fits to
the mass loss rates for M13, M15, and M92 are shown as given in Section 4.4 of the text. The two
coolest stars in M13 showed small outflow velocities and were not included
in the fitting procedure.
}
\end{figure}

\clearpage

\begin{deluxetable}{lccccccccc}
\tabletypesize{\scriptsize}
\tablecaption{Physical Parameters of Modeled Stars}
\tablewidth{0pt}
\tablehead{
\colhead{ID No. \tablenotemark{a}}           & \colhead{Cluster} &
\colhead{[Fe/H]}    &
\colhead{B$-$V}         &\colhead{V}    &
\colhead{T$_{\rm eff}$}     & \colhead{log~$L/L_{\odot}$}  & 
\colhead{$R/R_{\odot}$}&\colhead{log~g \tablenotemark{b}}  \\
\colhead{} &\colhead{} &\colhead{} &\colhead{} &\colhead{} &\colhead{(K)} &
\colhead{} &\colhead{} &\colhead{(cm s$^{2}$)} 
}
\startdata
L72	& M13 & $-$1.54 & 1.30 & 12.32 & 4180 & 3.096 & 65.7 & 0.71  \\
L96	& M13 & $-$1.54 & 1.27 & 12.52 & 4190 & 3.010 & 59.1 & 0.80  \\
L592	& M13 & $-$1.54 & 1.06 & 13.10 & 4460 & 2.689 & 36.0 & 1.23  \\
L954	& M13 & $-$1.54 & 1.54 & 12.09 & 3940 & 3.329 & 96.6 & 0.38  \\
L973	& M13 & $-$1.54 & 1.61 & 12.04 & 3910 & 3.377 & 103.0& 0.32  \\
K87	& M15 & $-$2.26 & 1.07 & 13.80 & 4610 & 2.708 & 34.5 & 1.27  \\
K341	& M15 & $-$2.26 & 1.37 & 12.81 & 4300 & 3.183 & 68.6 & 0.67  \\
K421	& M15 & $-$2.26 & 1.20 & 12.72 & 4330 & 3.207 & 69.5 & 0.66  \\
K479	& M15 & $-$2.26 & 1.31 & 12.68 & 4270 & 3.244 & 74.6 & 0.60  \\
K757	& M15 & $-$2.26 & 1.43 & 12.88 & 4190 & 3.195 & 73.2 & 0.61  \\
K969	& M15 & $-$2.26 & 1.16 & 13.45 & 4590 & 2.851 & 41.1 & 1.11  \\
VII-18	& M92 & $-$2.28 & 1.30 & 12.19 & 4190 & 3.208 & 76.3 & 0.41  \\
X-49	& M92 & $-$2.28 & 1.19 & 12.16 & 4280 & 3.184 & 71.2 & 0.48  \\
XII-8	& M92 & $-$2.28 & 1.06 & 12.76 & 4430 & 2.896 & 47.7 & 0.83  \\
XII-34	& M92 & $-$2.28 & 0.89 & 13.45 & 4660 & 2.570 & 29.6 & 1.24  \\
\enddata
\tablenotetext{a}{\citet{ludendorff01} is the identification for stars in M13, 
\citet{kustner01} is the identification for stars in M15, 
and \citet{sandage02} is the identification for stars in M92.}
\tablenotetext{b}{The gravity was calculated assuming M=0.8$M_{\odot}$ for each star.}
\end{deluxetable}

\clearpage

\begin{deluxetable}{lcccccc}
\tablecolumns{7}
\tabletypesize{\scriptsize}
\tablecaption{Physical Parameters of Calculated Chromospheric Models}
\tablewidth{0pt}
\tablehead{
\colhead{ID No.}	    &
\colhead{Obs\tablenotemark{a}} & \colhead{T$_{\rm max}$ \tablenotemark{b}} &
\colhead{$v_{\rm bis, 1}$}    & \colhead{$v_{\rm max}$ \tablenotemark{c}}	& 
\colhead{$v_{\rm esc}$ \tablenotemark{d}}	  & \colhead{MLR} \\
\colhead{} &\colhead{Date} & \colhead{(K)} & 
\colhead{{(km \ s$^{-1}$)}} & \colhead{{(km \ s$^{-1}$)}} & \colhead{{(km \ s$^{-1}$)}} & 
\colhead{($M_{\odot} yr^{-1}$)} 
}
\startdata
\cutinhead{M13}
L72	& 2 & 10300 & $-5.0 \ \pm$ 0.6 & 13.0 & 54.0 &   3.3e-09 \\
	& 5 & 10100 & $-5.7 \ \pm$ 0.8 & 9.0  & 54.0 &   2.3e-09   \\
L96	& 2 & 9900 & $-6.1 \ \pm$ 0.9 & 19.0 & 57.0 &   4.8e-09   \\
L592	& 5 & 8000 & $-2.6 \ \pm$ 0.3 & 8.5  & 69.0 &   2.6e-09   \\
L954	& 2 & 10000 & $-1.8 \ \pm$ 0.6 & 12.0 & 48.0 &   3.1e-09   \\
L973	& 2 & 10300 & $-1.9 \ \pm$ 0.4 & 6.5  & 46.0 &   1.6e-09   \\
\cutinhead{M15}					   
K87	& 8 & 8000 & $-3.9 \ \pm$ 1.3 & 9.0  & 65.0 &   1.4e-09   \\
K341	& 1 & 9900 & $-3.2 \ \pm$ 0.6 & 12.0 & 51.0 &   2.2e-09   \\
	& 6 & 10000 & $-6.9 \ \pm$ 1.0 & 13.0 & 51.0 &   2.4e-09   \\
	& 7 & 10300 & $-6.2 \ \pm$ 0.6 & 13.0 & 51.0 &   2.4e-09   \\
	& 8 & 10200 & $-6.3 \ \pm$ 0.9 & 13.0 & 51.0 &   1.7e-09   \\
K421	& 7 & 10250 & $-4.3 \ \pm$ 0.7 & 10.0 & 66.0 &   1.9e-09   \\
K479	& 8 & 10400 & $-0.7 \ \pm$ 0.7 & 12.0 & 64.0 &   2.3e-09   \\
K757	& 1 & 10200 & $-2.8 \ \pm$ 0.5 & 3.0  & 50.0 &   5.7e-10   \\
	& 6 & 10600 & $-8.9 \ \pm$ 1.1 & 16.0 & 50.0 &   3.0e-09   \\
K969	& 1 & 10500 & $-4.0 \ \pm$ 0.6 & 6.0  & 57.0 &   1.1e-09   \\
	& 8 & 10700 & $-1.7 \ \pm$ 0.3 & 9.0  & 57.0 &   1.7e-09   \\
\cutinhead{M92}					   
VII-18  & 3 & 10150 & $-3.0 \ \pm$ 1.0 & 15.0 & 49.0 &   2.0e-09   \\
	& 4 & 10200 & $-2.8 \ \pm$ 0.8 & 15.0 & 49.0 &   2.0e-09   \\
X-49	& 3 & 9950 & $-6.9 \ \pm$ 0.8 & 15.0 & 50.0 &   1.9e-09   \\
XII-8	& 3 & 9900 & $-5.6 \ \pm$ 0.8 & 11.0 & 57.0 &   2.0e-09   \\
XII-34  & 3 & 8000 & $-2.3 \ \pm$ 1.3 & 8.0  & 66.0 &   1.2e-09   \\
\enddata
\tablenotetext{a}{Observations: 1: 2005 May 22, 2: 2006 March 14, 3: 2006 May 7, 4: 2006 May 9, 
5: 2006 May 10, 6: 2006 May 11, 7: 2006 October 4, 8: 2006 October 7.}
\tablenotetext{b}{The maximum mass column density of all models is
  $1\times10^{-5}$ (g cm$^{-2}$) in 
the chromosphere, the stellar radius for each model is R=70R$_{\odot}$.}
\tablenotetext{c}{The maximum expansion velocity used in the models.}
\tablenotetext{d}{Escape velocity  calculated at the level with the highest expansion velocity assuming 
M=0.8M$_{\odot}$.}
\end{deluxetable}

\clearpage

\begin{deluxetable}{lccccc}
\tablecolumns{6}
\tabletypesize{\scriptsize}
\tablecaption{Mass Loss Rates (MLR) of Modeled Stars}
\tablewidth{0pt}
\tablehead{
\colhead{ID No.}	  & \colhead{MLR} & \colhead{MLR} & \colhead{MLR} & \colhead{MLR} & \colhead{MLR}\\
\colhead{} & 
\colhead{Average} & \colhead{Fit} &\colhead{Reimers\tablenotemark{a}}
&\colhead{SC\tablenotemark{b}} 
& \colhead{Origlia\tablenotemark{c}} \\
\colhead{} & 
\colhead{($M_{\odot} yr^{-1}$)} & \colhead{($M_{\odot} yr^{-1}$)} & \colhead{($M_{\odot} yr^{-1}$)}  &
\colhead{($M_{\odot} yr^{-1}$)} & \colhead{($M_{\odot} yr^{-1}$)} 
}
\startdata
\cutinhead{M13}
L72	&  2.8e-09 & 3.8e-09 & 4.1e-08 & 2.1e-08 & 4.0e-08 \\
L96	&  4.8e-09 & 3.6e-09 & 3.0e-08 & 1.4e-08 & 3.6e-08 \\
L592	&  2.6e-09 & 2.8e-09 & 8.8e-09 & 3.5e-09 & 2.2e-08  \\
L954	&  3.1e-09 & 4.6e-09 & 1.0e-07 & 7.1e-08 & 5.8e-08  \\
L973	&  1.6e-09 & 4.8e-09 & 1.2e-07 & 9.2e-08 & 6.2e-08 \\
\cutinhead{M15}
K87	&  1.4e-09 & 1.4e-09 & 8.8e-09 & 3.9e-09 & 2.2e-08  \\
K341	&  2.2e-09 & 2.0e-09 & 5.2e-08 & 3.2e-08 & 4.5e-08  \\
K421	&  1.9e-09 & 2.0e-09 & 5.6e-08 & 3.5e-08 & 4.6e-08 \\
K479	&  2.3e-09 & 2.1e-09 & 6.5e-08 & 4.3e-08 & 4.9e-08 \\
K757	&  1.8e-09 & 2.1e-09 & 5.7e-08 & 3.5e-08 & 4.6e-08 \\
K969	&  1.4e-09 & 1.5e-09 & 1.5e-08 & 7.1e-09 & 2.7e-08 \\
\cutinhead{M92}
VII-18  &  2.0e-09 & 2.1e-09 & 9.0e-08 & 4.8e-08 & 5.5e-08 \\
X-49	&  1.9e-09 & 2.0e-09 & 7.8e-08 & 4.2e-08 & 5.2e-08   \\
XII-8	&  2.0e-09 & 1.7e-09 & 2.7e-08 & 1.0e-08 & 3.4e-08  \\
XII-34  &  1.2e-09 & 1.4e-09 & 7.9e-09 & 2.5e-09 & 2.1e-08  \\
\enddata
\tablenotetext{a}{Rate from \citet{reimers01,reimers02}.} 
\tablenotetext{b}{Rate from \citet{schroder01}.}
\tablenotetext{c}{Rate from \citet{origlia02}.}
\end{deluxetable}

\end{document}